\documentclass[referee]{ati}		
\usepackage{graphicx,times}
\usepackage{lmodern}
\usepackage{natbib}
\usepackage{url}
\usepackage{caption}
\usepackage{rotating}
\usepackage[colorlinks=true,breaklinks=true, linkcolor=red,urlcolor=magenta,citecolor=blue,anchorcolor=green]{hyperref}

\begin{document}
   \title{The Jiao Tong University Spectroscopic Telescope Project}


   \author{JUST Team  \correspondingAuthor{}
      \inst{1,2}
    \and Chengze Liu\inst{2,1}
    \and Ying Zu\inst{2,1}
    \and Fabo Feng\inst{1,2}
    \and Zhaoyu Li\inst{2}
    \and Yu Yu\inst{2}
    \and Hua Bai\inst{3,4}
    \and Xiangqun Cui\inst{3,5}
    \and Bozhong Gu\inst{3}
    \and Yizhou Gu\inst{1,2}
    \and Jiaxin Han\inst{2}
    \and Yonghui Hou\inst{3,5}
    \and Zhongwen Hu\inst{3,5}
    \and Hangxin Ji\inst{3}
    \and Yipeng Jing\inst{1,2}
    \and Wei Li\inst{6}
    \and Zhaoxiang Qi\inst{7}
    \and Xianyu Tan\inst{1}
    \and Cairang Tian\inst{6}
    \and Dehua Yang\inst{3}
    \and Xiangyan Yuan\inst{3,4}
    \and Chao Zhai\inst{8}
    \and Congcong Zhang\inst{7}
    \and Jun Zhang\inst{2}
    \and Haotong Zhang\inst{9}
    \and Pengjie Zhang\inst{1,2}
    \and Yong Zhang\inst{9}
    \and Yi Zhao\inst{6}
    \and Xianzhong Zheng\inst{10}
    \and Qingfeng Zhu\inst{8}
    \and Xiaohu Yang\inst{1,2}
   }
\correspondent{JUST Team}	
\correspondentEmail{just.astro@sjtu.edu.cn}
\institute{Tsung-Dao Lee Institute, Shanghai Jiao Tong University, Shanghai 200240, China; 
        \and Department of Astronomy, School of Physics and Astronomy, Shanghai Jiao Tong University, Shanghai 200240, China; 
        \and Nanjing Institute of Astronomical Optics \& Technology, Chinese Academy of Sciences, Nanjing 210042, China; 
        \and University of Chinese Academy of Sciences, Nanjing 211135, China; 
        \and University of Chinese Academy of Sciences, Beijing 101408, China; 
        \and Lenghu Technology Innovation Industrial Park Management Committee, Lenghu 817400, China; 
        \and Shanghai Astronomical Observatory, Chinese Academy of Sciences, Shanghai 200030, China; 
        \and Department of Astronomy, University of Science and Technology of China, Hefei 230026, China; 
        \and National Astronomical Observatories, Chinese Academy of Sciences, Beijing 100012, China; 
        \and Purple Mountain Observatory, Chinese Academy of Sciences, Nanjing 210023, China. 
}

   \date{Received~~2023 month day; accepted~~2023~~month day} 
   \abstract{
The Jiao Tong University Spectroscopic Telescope (JUST) is a 4.4-meter $f/6.0$ segmented-mirror telescope
dedicated to spectroscopic observations. The JUST primary mirror is composed of 18 hexagonal segments, each with a
diameter of 1.1 m. JUST provides two Nasmyth platforms for placing science instruments. One Nasmyth focus fits a
field of view of $10^\prime$  and the other has an extended field of view of $1.2^\circ$ with correction optics. A tertiary mirror is
used to switch between the two Nasmyth foci. JUST will be installed at a site at Lenghu in Qinghai Province, China,
and will conduct spectroscopic observations with three types of instruments to explore the dark universe, trace the
dynamic universe, and search for exoplanets: (1) a multi-fiber (2000 fibers) medium-resolution spectrometer (R=4000-5000)
to spectroscopically map galaxies and large-scale structure; (2) an integral field unit (IFU) array of 500 optical fibers
and/or a long-slit spectrograph dedicated to fast follow-ups of transient sources for multi-messenger astronomy; (3) a
high-resolution spectrometer (R$\sim$100000) designed to identify Jupiter analogs and Earth-like planets, with the capability
to characterize the atmospheres of hot exoplanets.
}

   \authorrunning{ASTRONOMICAL TECHNIQUES \& INSTRUMENTS }   
   \titlerunning{JUST Team:~ (JUST) Project}  
   \VolumeNumberPageYear{1}{1}{1}{2030} 
   \MonthIssue{Jan}			
   \DOItail{0000000.000}	   	
   \maketitle
\setcounter{page}{\Page}	
%
%
\section{Introduction}
\label{sect:intro}

Observing facilities play a fundamental role in advancing our understanding of the universe. These facilities,
including ground-based telescopes, space observatories,
and specialized instruments, provide astronomers the necessary tools to gather data from distant celestial objects and
phenomena, and explore the properties, compositions, and
behaviors of objects such as stars and galaxies, leading to
remarkable discoveries and profound insights into the
nature of the universe. Moreover, long-term observations 
with these facilities enable monitoring of transient events,
and probe the cosmos across various wavelengths, which
is essential for unveiling cosmic mysteries. Observing facilities are indispensable for pushing the boundaries of astronomical knowledge and fostering scientific breakthroughs.
The development of powerful observing facilities, whether
for general-purpose use or dedicated surveys, has become
a critical requirement for astronomers to achieve groundbreaking advancements. The progress of astronomy hinges
on the construction of large telescopes which are currently evolving to possess large apertures, wide fields of
view, and high spatial and spectral resolution. Given the disparity in associated cost between acquiring images and spectra, there is a notable shortfall in high-quality spectroscopic observational facilities compared with the plentiful availability of image-based observational facilities.
This gap highlights the need for further attention and investment in advancing spectroscopic capabilities to complement the existing observational landscape.

Astronomical spectroscopy enables the precise measurement of redshift, identification of specific chemical
elements, and the determination of kinematics of celestial
objects. It leads to a deeper understanding of the nature
and characteristics of the observed objects. Spectroscopic
observations offer a wealth of information that complements and enhances the insights gained from imaging observations.

\begin{figure*}[htb!]
    \begin{minipage}[t]{0.999\linewidth}  
    \centering
    \includegraphics[width=1.0\textwidth]{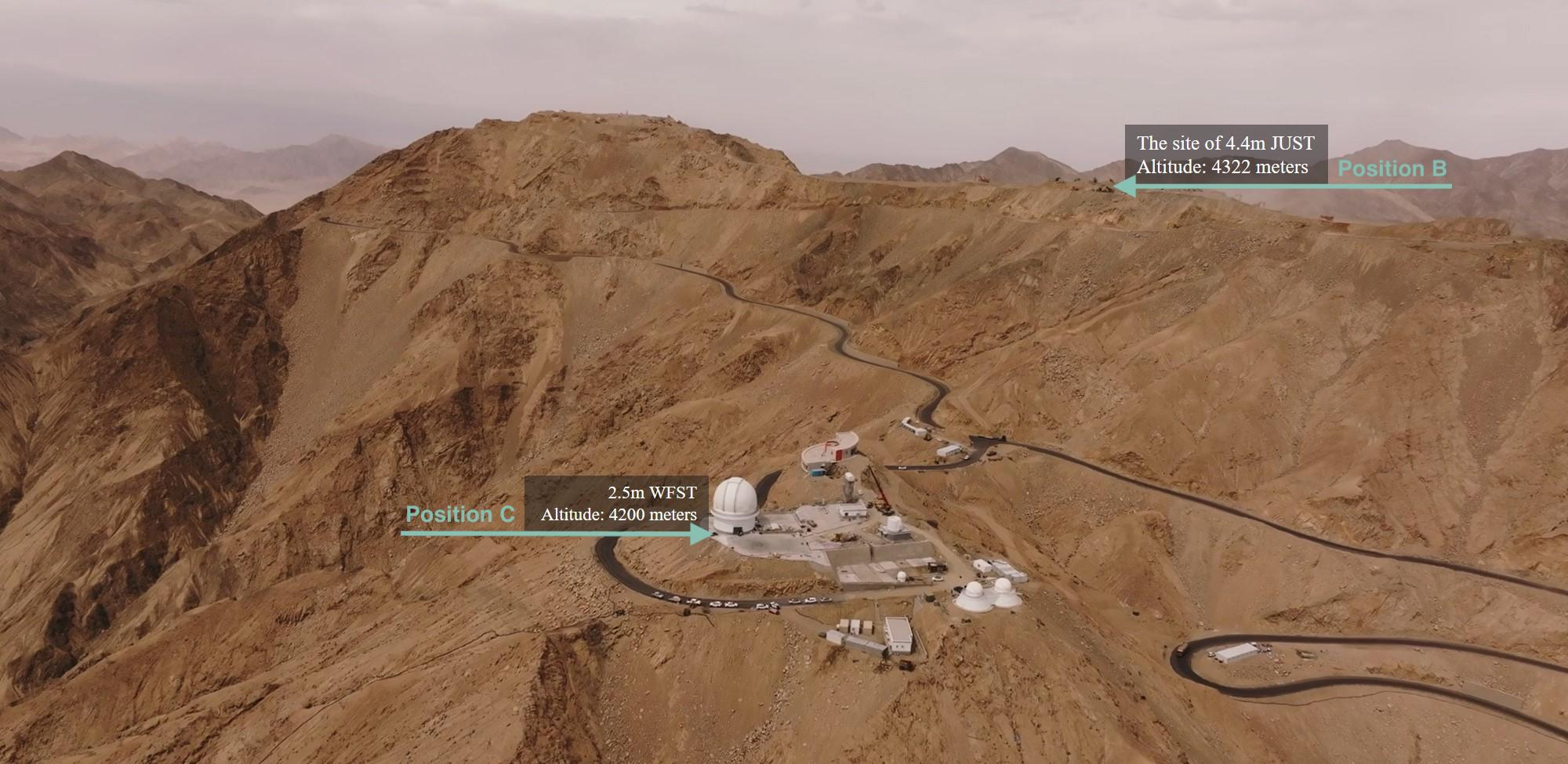} 
    \caption{Bird's-eye view of Saishiteng Mountain. The largest dome at Position C (at an altitude of 4200 \,m) houses the WFST,
which is dedicated to imaging surveys. JUST will be placed at Position B (at 4322 \,m). Photo credit: Bin Chen.}
    \label{fig:site}
    \end{minipage}
\end{figure*}

To fulfill the scientific needs for spectroscopic observations, as well as owing to the great success of spectroscopic projects such as the Sloan Digital Sky Survey \citep[SDSS\footnote{\url{https://sdss.org/}};][]{2000AJ....120.1579Y},
multiple new projects are thriving. The Dark Energy Spectroscopic Instrument \citep[DESI\footnote{\url{https://www.desi.lbl.gov}};][]{2016arXiv161100036D} is the first stage-IV dark
energy survey project, comprising a 4-meter telescope with 5 000 robotic fiber positioners to feed a collection of spectrographs covering the 360-980 nm wavelength
range. It has reportedly finished over 50
before the planned 5 years of run time, demonstrating its
high efficiency in observation. Near future 4-meter-class
telescope projects include the WHT Enhanced Area Velocity Explorer \citep[WEAVE;][]{2023MNRAS.tmp..715J} and the 4-metre Multi-Object
Spectroscopic Telescope \citep[4MOST\footnote{\url{https://www.4most.eu}};][]{2019Msngr.175....3D}.
They will provide the spectroscopic follow-up
required for full scientific exploitation of other projects,
such as the Gaia, LOFAR and Apertif surveys. 
The MegaMapper \citep{2019BAAS...51g.229S} will be a dedicated cosmology facility with highly efficient redshift measurements on a 6.5 m telescope. 8-meter-class projects include the Subaru Prime Focus Spectrograph \citep[PSF\footnote{\url{https://pfs.ipmu.jp}};][]{2022SPIE12184E..10T} project, and the Multi-Object Optical and Near-IR Spectrograph \citep[MOONS\footnote{\url{https://vltmoons.org}};][]{2020Msngr.180...10C}.
With increasing telescope size, the Maunakea Spectroscopic Explorer \citep[MSE\footnote{\url{https://mse.cfht.hawaii.edu}};][]{2018arXiv181008695H}, SpecTel \citep{2019BAAS...51g..45E} and 
the Fiber-Optic Broadband Optical Spectrograph  \citep[FOBOS\footnote{\url{https://fobos.ucolick.org/}};][]{2019BAAS...51g.198B} are 10-meter-class projects.
All of these large telescopes will equip instruments with thousands to tens of thousands of optical fibers, aiming at simultaneous spectroscopic observations of multiple objects at once.

In China, optical telescopes currently fall behind
world-class standards. However, with improved funding
availability and technological capabilities, observatories
and universities have initiated the construction of optical
telescopes with diameters exceeding 2 meters. This initiative is driven by diverse scientific objectives, and aims to
facilitate distinctive observational research, enabling universities within China to make substantial progress with medium-sized observing facilities, such as the Wide Field Survey Telescope (WFST or ``Mocius''\footnote{\url{https://wfst.ustc.edu.cn}}) is one of them and is on commissioning phase \citep{2023SCPMA..6609512W}. For spectroscopic observations, there are several telescopes, either proposed or under construction, such as the 4.4-meter Jiao Tong University Spectroscopic Telescope (JUST\footnote{\url{https://just.sjtu.edu.cn}}), and 6.5-meter
MUltiplexed Survey Telescope (MUST\footnote{\url{https://must.astro.tsinghua.edu.cn}}), as well as the stage II  of Large Sky Area Multi-Object Fiber Spectroscopic Telescope (LAMOST-II\footnote{\url{https://www.lamost.org}}). 

Construction has already begun on JUST, and the
first light observations are expected within three years.
This telescope will be equipped with dedicated spectroscopic instruments to explore the dark universe, trace the
dynamic universe, and search for exoplanets. In this
paper, we provide basic information about details such as
the site, structure, optical system, and science motivations of JUST. The structure of this paper is outlined as follows. We first introduce the site condition in Section \ref{sec:site}, followed by the conceptual design of the JUST telescope in
Section \ref{sec:design}. An overview of the planned instruments and science motivations is presented in Section \ref{sec:instruments}. We summarize the JUST project in Section \ref{sec:summary}. 

\section{Site and dome} \label{sec:site}
Mauna Kea in Hawaii and certain summits and
plateaus in northern Chile are among the best observing
sites on the Earth. Over the past two decades, much effort
has been dedicated to the search for excellent astronomical sites in China. Recently, the summit of Saishiteng
Mountain in Lenghu, located on the Tibetan Plateau, was
identified as possessing favorable observing conditions.
Site monitoring has shown that the summit of Saishiteng
Mountain, situated at an altitude of 4 200 to 4 500
meters, experiences clear nights for approximately 70\% of
the year and boasts good median seeing of 0.75 \,arcsec \citep{2021Natur.596..353D}. The climate in the surrounding area of the site is extremely arid, and the sky background is exceptionally
dark due to minimal light pollution. Furthermore, the time
zone of the Lenghu site is distinct from that of nearly all
observatories worldwide, facilitating complementary timedomain astronomy.

The planned installation of the JUST telescope is
shown in Fig.~\ref{fig:site}, at position B, situated at an altitude of
4322 m on the summit of Saishiteng Mountain, close to
position C where WFST is located. Spectroscopic observations with JUST will complement imaging observations
made with WFST, providing essential photometric and spectroscopic data for advancing researches across various
domains of astronomy.

\begin{figure}[htb!]
    \begin{minipage}[t]{0.999\linewidth}  
    \centering
    \includegraphics[width=1.0\textwidth]{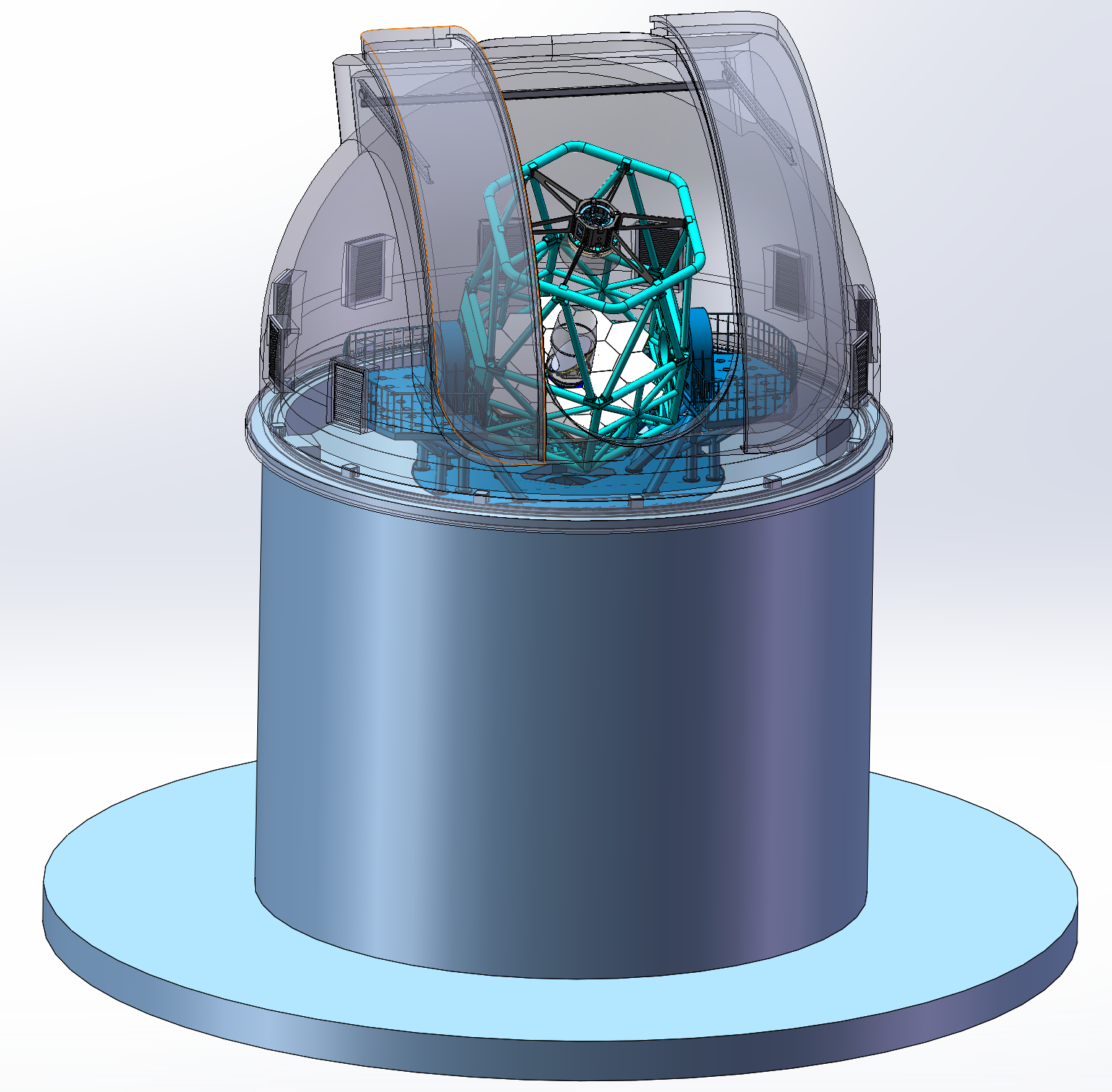} 
    \caption{The conceptual design of the dome for JUST.}
    \label{fig:dome}
    \end{minipage}
\end{figure}

Fig.~\ref{fig:dome} illustrates the design concept of the telescope
dome for JUST. It incorporates a classical semi-spherical
dome, featuring a shutter that can be opened to allow the
telescope to observe. The dome will also integrate ventilation systems to regulate temperature and air circulation,
improving dome seeing. Additionally, it will include lighting and other equipments to support telescope operation
and maintenance. The construction of the site infrastructure and telescope dome is scheduled to commence in
2024. 

\section{JUST main part conceptual design} \label{sec:design}
\subsection{General parameters}

\begin{figure*}[ht]
    \begin{minipage}[t]{0.999\linewidth}  
    \centering
    \includegraphics[width=0.7\textwidth]{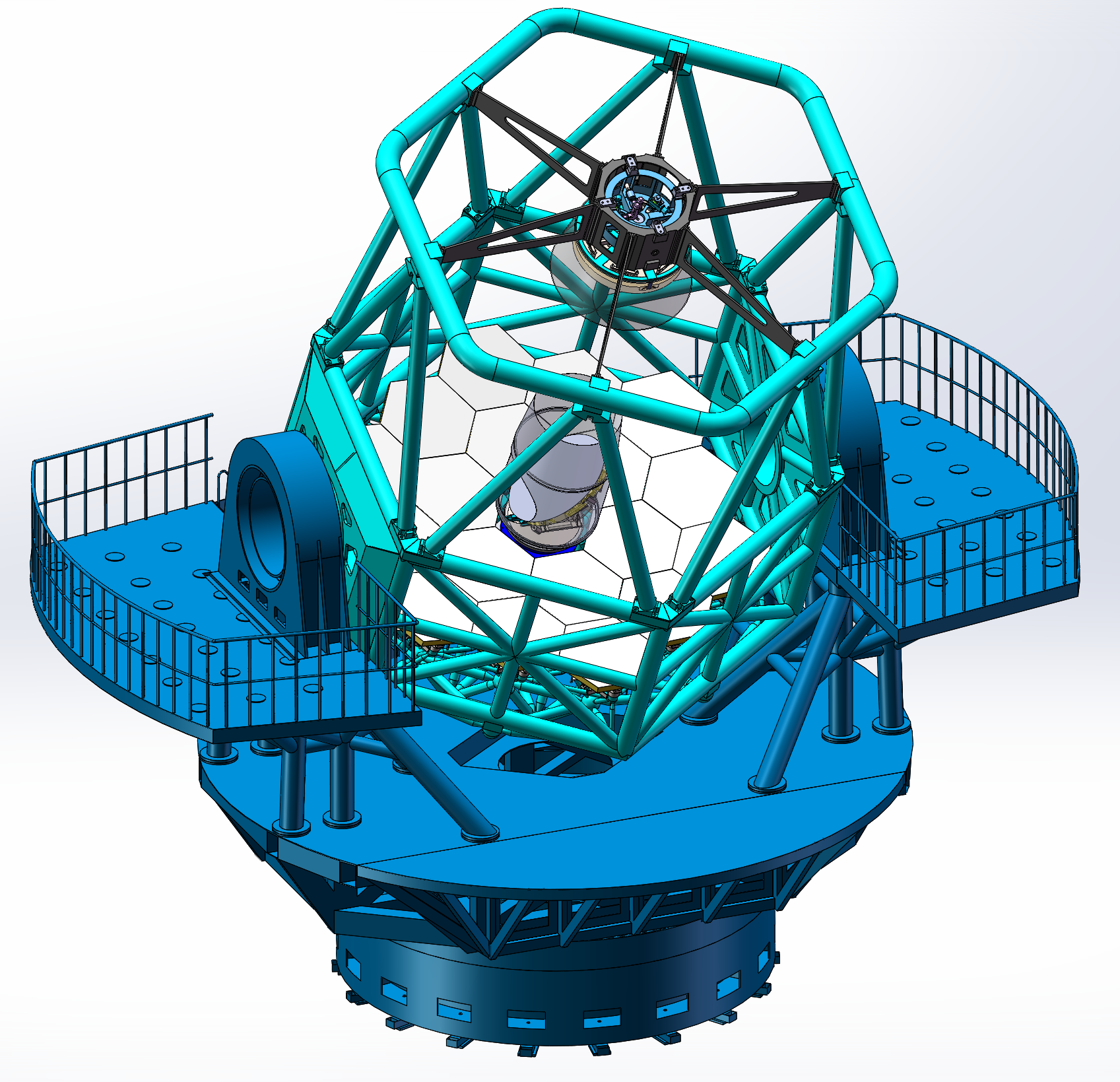} 
    \caption{The conceptual design of the structure of JUST.}
    \label{fig:structure}
    \end{minipage}
\end{figure*}

The telescope project encompasses three main functional subsystems: telescope optics, support and structure,
and telescope control. The telescope optics subsystem comprises optical mirrors and mirror support with active
optics. The support and structure subsystem includes a
tracking mount and telescope tube. The control subsystem incorporates the telescope control subsystem (TCS),
observation control subsystem, and active optics control
subsystem. A lightweight telescope design is achieved
through the selection of a horizontal tracking mount and a
truss-type telescope tube structure. The overall conceptual view of the telescope is illustrated in Fig.~\ref{fig:structure}.

\begin{figure*}[ht!]
\includegraphics[width=\textwidth]{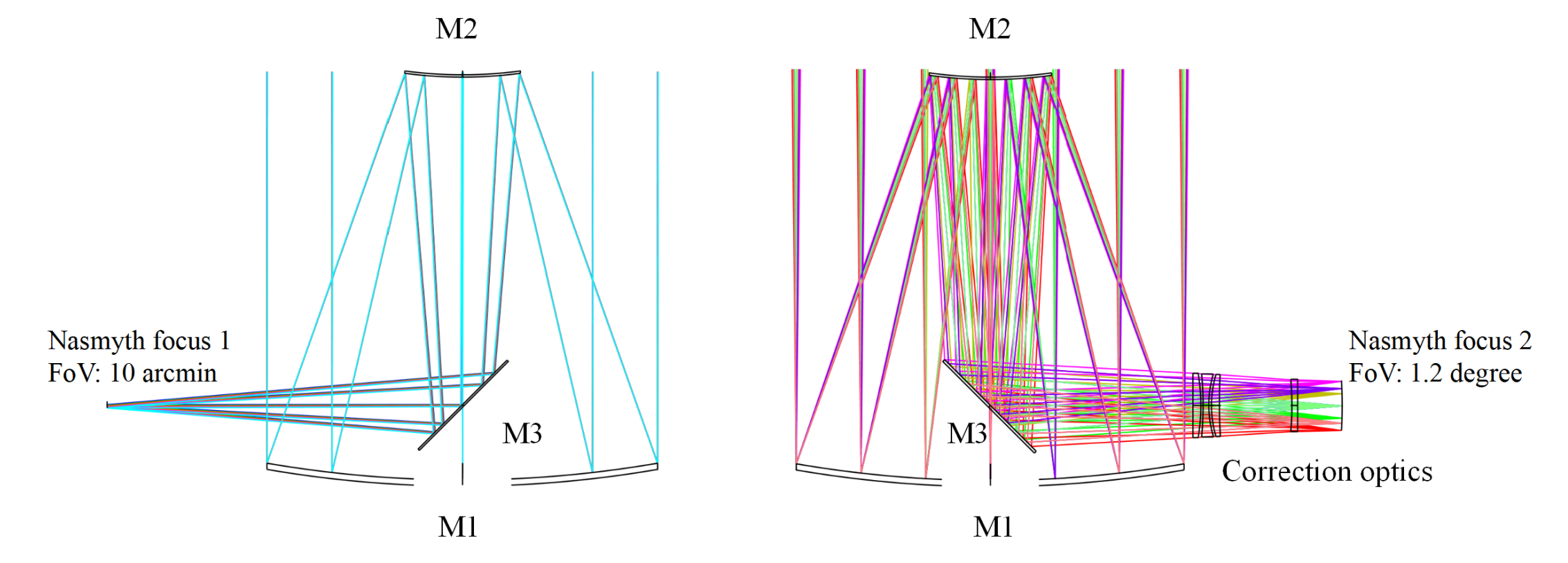}
\caption{Optical design of JUST. \textit{Left:} Nasmyth focus 1 with field of view of $10^\prime$; \textit{Right:} Nasmyth focus 2 with an extended field of view of $1.2^\circ$. 
\label{fig:optics}}
\end{figure*}

JUST has two Nasmyth foci with a focal ratio of
$f/6.0$. One Nasmyth focus has a field of view (FoV) of
$10^{\prime}$ and the other Nasmyth focus has an extended FoV of
$1.2^\circ$ with correction optics, as shown in Fig.~\ref{fig:optics}. 
The two Nasmyth foci can be switched by rotating the tertiary mirror (M3). 

Nasmyth focus 1 is a purely reflecting system,
in which high-resolution wide-wavelength instruments or
infrared instruments can be mounted. The image quality is
defined as the full width at half maximum (FWHM) of
the image profile. The intended image quality of Nasmyth focus 1 is $0.09^{\prime\prime}$. With manufacturing, alignment, and control error, this value can be reduced to $0.35^{\prime\prime}$.

Nasmyth focus 2 is used for the Multi-Object Fiber
Spectroscopic Survey. The diameter of the focal plane is
570 mm, and about 2 000 optical fibers can be accommodated. At a zenith distance of $60^\circ$, observation site altitude of 4 200 m, and the wavelength range of 0.35-1.3 $0.35-1.3$ $\mu$m, the atmospheric dispersion is $3.1^{\prime\prime}$. 
However, astigmatism rapidly increases with the square of the FoV, so it
is essential to include a corrector for widening FoV and
compensating for atmospheric dispersion. The corrector consists of four silica lenses, two of which are the atmospheric dispersion correctors (ADCs). The target image
quality of Nasmyth focus 2 is $0.51^{\prime\prime}$. With errors, the delivered image quality will be $0.7^{\prime\prime}$, which is close to the
value of median seeing at the Lenghu site. As a reference, we list in Table~\ref{tab:optics} the main optical parameters of JUST. 

\begin{table}  
    \begin{minipage}[t]{0.999\linewidth}  
    \caption[ ]{Optical parameters.}
    \label{tab:optics} 
    \end{minipage}
    \begin{center}
    \begin{tabular}{ll}
    \hline\hline\noalign{\smallskip}
    \multicolumn{1}{c}{Parameters}  & \multicolumn{1}{c}{Values}  \\
    \hline\noalign{\smallskip}
 primary mirror diameter & 4.4\,m (segmented) \\
 M1 focal length & $6.4$\,m \\
 System $F$ ratio  & $6.0$ \\
 Focal scale of Focus 1 \& 2 & $7.8$\,arcsec\,mm$^{-1}$ \\
 Focus 1 & Nasmyth focus \\
         & (high precision) \\
 Field of view 1  & $10$\,arcmin \\
 Wavelength range 1 & Purely reflecting \\
 Focus 2 & Nasmyth focus \\
         & (large field of view) \\
 Field of view 2  & $1.2$\,degree  \\
 Wavelength range 2 & $0.35-1.3\,\mu$m \\
   \hline\noalign{\smallskip}
   \end{tabular}
   \end{center}
\end{table}

\subsection{The mirrors} \label{sec:mirrors} 
\begin{figure*}[ht!]
\centering
\includegraphics[width=0.5\textwidth]{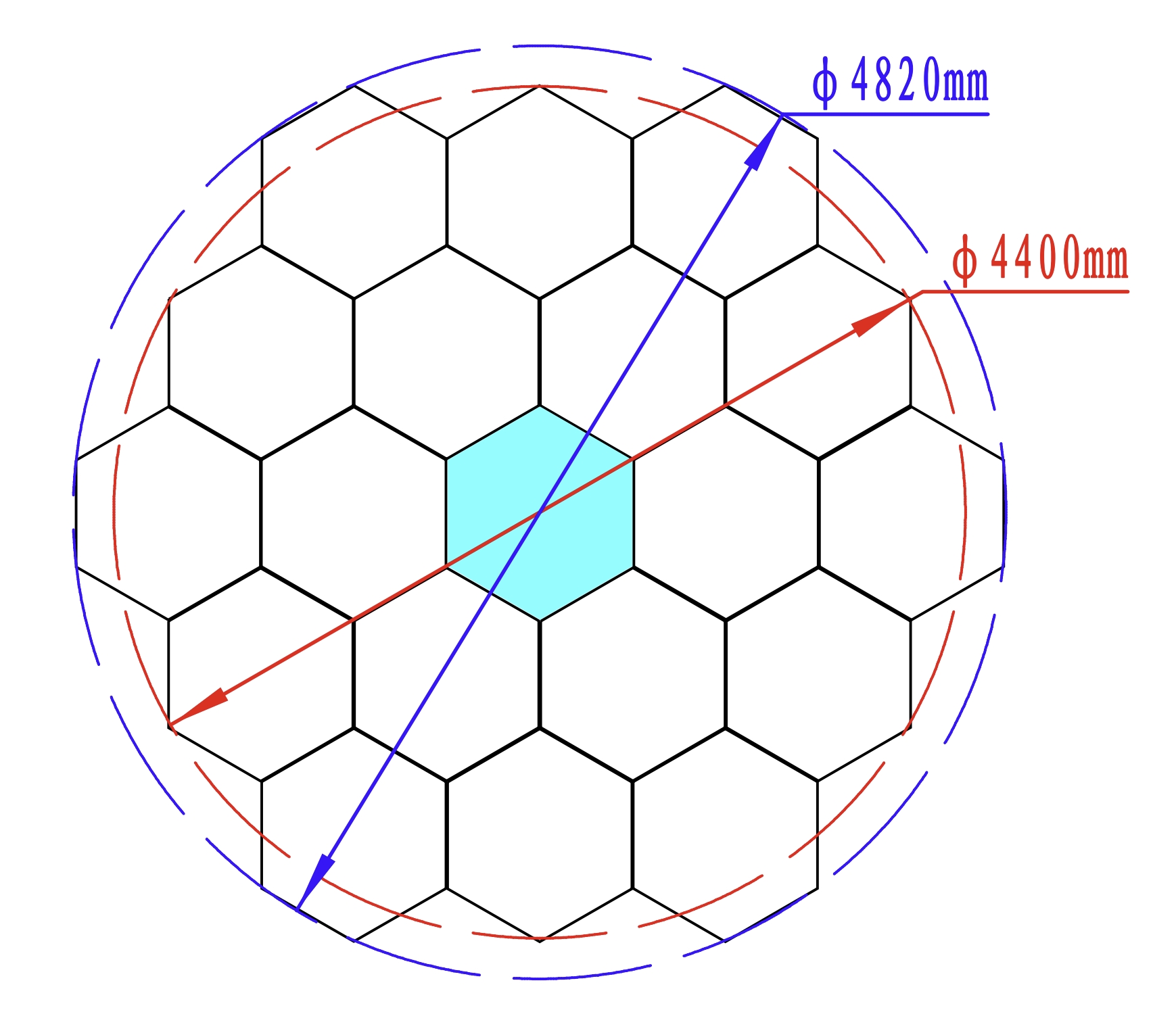}
\caption{Configuration of the segments of M1, showing 18
hexagonal segments. The central area of the primary mirror is
vacant, where M3 will be installed.  
\label{fig:M1}}
\end{figure*}

The primary mirror (M1) is composed of 18 hexagonal segments, with an effective aperture of 4.4 m. The detailed configuration is depicted in Fig.~\ref{fig:M1}. Each segment
is equipped with its own support system to maintain the
correct optical surface. The axial support system utilizes
an 18-point whiffletree support for the segments, while
the radial support employs a central flexible support. This
support system serves the following functions: 
\begin{itemize}
\item Accurate installation of the segments onto the main
truss; 
\item Support for the segments to meet the requirements
for the mirror surface shape;
\item Active optical technology to control the closed-loop
segmented system of the primary mirror, mitigating the
effects of temperature and gravity, and achieving co-focusing/co-phasing of the segments. 
\end{itemize}

The secondary mirror (M2) support module is
designed to preserve its original machining accuracy and
stabilize its spatial position. The support system for the secondary mirror includes a bottom support, lateral support,
and centering mechanism. The bottom support features a
suspended whiffletree support structure, while the lateral
support uses a lever-balanced weight support structure.
The centering mechanism employs a bi-directional membrane structure in both radial and axial directions. 

The tertiary mirror (M3) uses a whiffletree floating support structure, with the centering mechanism also adopting a bi-directional membrane structure to serve the radial
and axial directions, in addition to function as lateral support. 

\subsection{Active optics} \label{sec:active_optics}

The M1 is designed to use active optics technology
for real-time closed-loop control splicing to join the segments into a single mirror surface. The active optical system primarily comprises core devices such as segment surface support, displacement actuator, and an active optical
wavefront sensor. 

\begin{itemize}
\item Segment surface support system: Ensures that the support surface meets and exceeds the technical requirements for the segment surface shape. 
\item Displacement actuator: Utilizes nano-electromechanical displacement actuators controlled in parallel by multiple active optical intelligent controllers to achieve nanometer-level displacement resolution and millimeter-level displacement range output accuracy and stroke under the full load of the segments.
\item Active optical wavefront sensor: Uses a Shack-Hartmann wavefront sensor based on physical optics. It measures the surface shape, imaging quality of individual segments, and the segmented primary mirror. It uses the central star as a target source, providing precise feedback to
continuously drive the displacement actuators for segmented mirror calibration and maintenance. 
\end{itemize}

Upon implementation of active optics technology, the
telescope will achieve co-focusing of the primary mirror
and obtain imaging quality close to the visibility limit of
the telescope's location, in conjunction with the optical system design. 

\begin{table}  
    \begin{minipage}[t]{0.999\linewidth}  
    \caption[ ]{Key parameters of three types of spectrographs.} 
    \label{tab:instrument}
    \end{minipage}
    \begin{center}
    \begin{tabular}{lll}
    \hline\hline\noalign{\smallskip}
    \multicolumn{1}{c}{Parameters}  & \multicolumn{1}{c}{fibers} & \multicolumn{1}{c}{Resolution}  \\
    \hline\noalign{\smallskip}
 multi-object spectrograph & 2~000 & $4~000-5~000$ \\
 IFU  array & 500 & $4~000-5~000$ \\
 long-slit spectrograph & N/A & $4~000-5~000$\\
high-resolution spectrograph & 1-3 & $\sim 100~000$ \\
   \hline\noalign{\smallskip}
   \end{tabular}
   \end{center}
\end{table}

\section{Science cases and scientific instruments} \label{sec:instruments}
JUST has two Nasmyth platforms and will be
installed with three types of spectrographs on both. The
basic parameters of these spectrographs are listed in
Table~\ref{tab:instrument}. 
\begin{itemize}
    \item Galaxies and large-scale structures: JUST will be
equipped with multiple fiber positioners and medium-resolution spectrometers to conduct spectral surveys of a large
number of galaxies. 
    \item Multi-messenger astronomy: JUST will be equipped
with hundreds of optical fibers to form an Integrated
Field Unit (IFU) array and/or a long-slit spectrograph for
follow-up observation of a large number of transient
sources. 
    \item Exoplanet detection and characterization: JUST will
use advanced high-resolution spectrometers to detect cold
giant planets and earth-like terrestrial planets, and to provide detailed atmospheric characterization for hot exoplanets. 
\end{itemize}

\subsection{Exploring  the dark Universe}


More than 95\% of the Universe remains dark to humanity, whether in the form of dark matter or dark energy~\citep{Planck2020}. The first step toward understanding the dark Universe requires the accurate measurement of the growth of cosmic structures on scales ranging from a few kiloparsecs to hundreds of megaparsecs, with highly-multiplexed spectroscopic surveys of galaxies~\citep{Weinberg2013}. To complement the current stage-IV surveys that focus on the galaxy distribution at linear scales, JUST will dedicate its multi-object spectroscopic~(MOS) capability to the mapping of structures from quasi-linear to highly non-linear scales, centered on massive galaxy clusters at $z{<}0.6$. JUST aims for complete spectroscopic coverage of the galaxies at $r<20$ mag in the cosmic web surrounding clusters below $z<0.6$, complementing DESI spectra.

The JUST spectroscopic cluster survey(SCS) will improve cluster cosmology as one of the most sensitive probes of cosmic growth, through the mitigation of systematic uncertainties in the cluster redshifts, satellite membership assignment, and various projection effects associated with photometric cluster finders~\citep{Erickson2011, Noh2012, Zu2017, Costanzi2019}. The spectroscopic cluster catalogue will provide stringent constraints of key cosmological parameters, including the matter density, the amplitude of matter clustering, the equation-of-state of dark energy, and the sum of neutrino masses~\citep{Sartoris2016}. The combination of the redshift-space distortion(RSD) of infalling galaxies~\citep{Lam2013, Zu2013, Hamabata2019, Shirasaki2021}
and the weak lensing of background sources~\citep{Johnston2007, Simet2017, Wang2022} by galaxy clusters will enable stringent tests of theories of cosmic acceleration and distinguish between dark energy and modified gravity on inter-cluster scales~\citep{Zu2014, Koyama2016, Joyce2016, Baker2021}. Meanwhile, JUST-SCS will fully sample cluster galaxies in both the velocity phase space(cluster-centric radius vs. line-of-sight velocity) and the color-magnitude diagram, from infall to the splashback regions, and into the virialized cores of clusters~\citep{FG1984, Bertschinger1985, Kravtsov2012, Diemer2014, More2016, Walker2019}. Such spectroscopic coverage of the cosmic web will provide a comprehensive picture of galaxy formation in different environments surrounding galaxy clusters~\citep{Kauffmann2004}. 

\begin{figure*}[ht!]
    \centering
\includegraphics[width=0.7\textwidth]{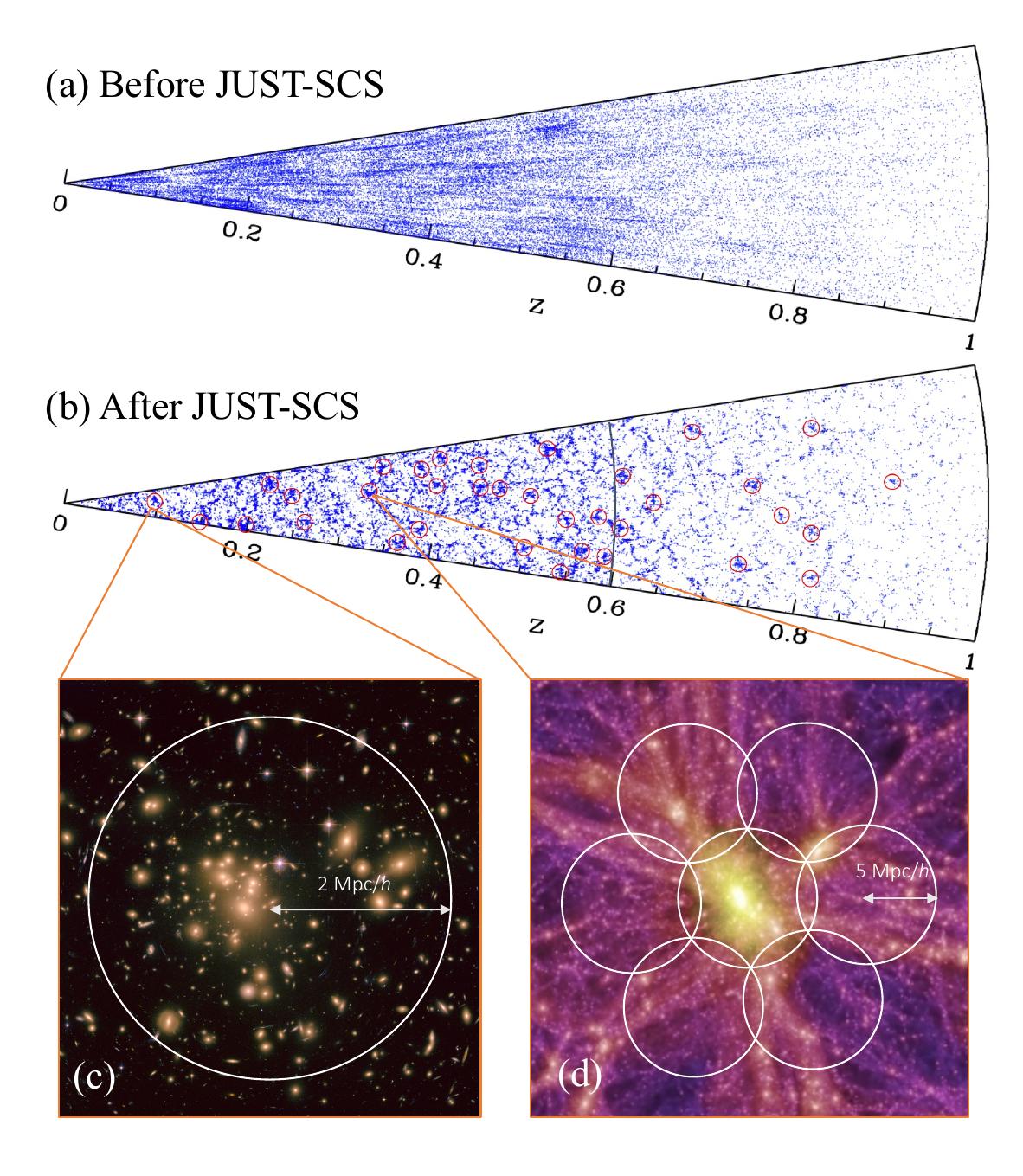}
\caption{Illustration of the JUST-SCS program. \textit{Panel(a):} Distribution of the photometric galaxies within a simulated lightcone. Galaxies from the same cluster are dispersed over a large line-of-sight distance due to photo-z uncertainties.  \textit{Panel(b):} Distribution of the spectroscopic clusters~(red circles) that will be observed by the JUST cluster cosmology survey within the same lightone. \textit{Panel(c):} A cluster at $z=0.1$ within the JUST field of view~(white circle) targeted by the JUST cluster galaxy evolution survey. The Background image is Abell 1689. 
\textit{Panel(d):} The cosmic web structure centered on a cluster at $z{\simeq}0.3$ targeted by the JUST cluster infall survey. The background image is from the Millennium Simulation. 
\label{fig:justscs}}
\end{figure*}

In recent decades, Chinese astronomers have made significant contributions to revealing the nature of the dark universe with contributions such as measuring and quantifying large scale structure, elucidating the galaxy-halo connection, constraining the cosmological parameters. Among these efforts, representative work includes establishing the halo occupation distribution model~\citep{Jing1998} based on the Las Campanas Redshift Survey \citep{Shectman1996}, establishing the conditional luminosity function model \citep{Yang2003} based on the 2-degree Field Galaxy Redshift Survey(2dFGRS) \citep{Colless2001}, establishing the halo-based group finder \citep{Yang2005, Yang2007} based on 2dFGRS and the Sloan Digital Sky Survey \citep{York2000}, and making dark energy model constraints \citep{Zhao2017} based on the Baryon Oscillation Spectroscopic Survey \citep{Alam2015}. Most of these achievements were made based on either public data releases or through international collaborations of large galaxy redshift surveys. With JUST-SCS, we will have greater opportunity to explore the dark universe with our own observational data set. To maximize the science return of the MOS survey on cluster cosmology and galaxy evolution, JUST-SCS will include three layers as summarized by Figure~\ref{fig:justscs}. We will discuss each of the three in the subsections below.

\subsubsection{JUST Cluster Cosmology Survey}

The upcoming China Space Survey Telescope \citep[{\it CSST};][]{Miao2023}) will detect approximately 300,000 photometric halo-based cluster candidates with halo mass above $10^{14}\,M_\odot/h$ up to $z<1.5$ \citep{Yang2021}, serving as the basis of target selection for JUST-SCS. In particular, the JUST cluster cosmology survey will target ${\simeq}50,000$ clusters over $10,000$ deg$^2$ at $z{<}0.6$, producing an unprecedented spectroscopic cluster sample for cosmological analysis. For each cluster, JUST will be used to obtain spectra for the brightest cluster galaxy~(BCG) and the bright member galaxy candidates down to $r{=}20$, including but without re-observing the spectra from the full DESI survey. This program will not only provide secure spectroscopic redshifts for a cosmologically significant volume of individual clusters, but also improve the centering of clusters both perpendicular and along the line-of-sight~\citep{Sohn2021}. Such an accurate localization of individual clusters in three dimensions enables cosmological analyses using massive dark matter haloes, instead of galaxies, as spectroscopic tracers of the large-scale structure. 

With spectroscopic redshifts for up to 20 member galaxy candidates, JUST will be able to disentangle the chance alignment of structures along the line of sight, and mitigate interlopers from any correlated structures on the velocity phase diagram. The spectroscopically confirmed satellite galaxies will enable mass estimates of individual haloes through the velocity dispersion~\citep{Evrard2008, Wu2013, Ntampaka2015} and caustic boundary ~\citep{Diaferio1999, Gifford2013, Rines2013}, improving the calibration of the cluster mass-observable relation beyond the optical richness~\citep{Rozo2009}. Meanwhile, JUST will probe the properties of the intracluster medium, particularly the circumgalactic medium of cluster galaxies, by measuring metal absorption lines recorded in the DESI spectra of background quasars in the cluster fields~\citep{Zhu2013, Lee2021, Zu2021, Anand2022, Napolitano2023}.

\subsubsection{JUST Cluster Infall Survey}

In the intermediate redshift range between $0.1{<}z{<}0.4$, JUST-SCS aims to achieve a complete spectroscopic coverage of galaxies within an approximately 20 Mpc/h radius surrounding each cluster down to r = 20, on top of the existing spectra from the DESI Bright Galaxy Survey~\citep[BGS;][]{Hahn2023}. In addition, JUST will spectroscopically cover a large number of non-cluster fields to the same depth, as the control sample of field galaxies for the cluster-galaxy cross-correlation measurements and the galaxy evolution study. The target selection of the non-cluster fields will be optimized based on the signal-to-noise forecast of the cluster RSD analysis.

The JUST cluster infall survey will push the($\texttt{E}_G$) method \citep{Zhang2007} from the linear regime to the infall region around clusters, where the potential imprint of modified gravity remains unscreened and the signal-to-noise of the RSD and weak lensing measurements is high. In particular, JUST will accurately measure the cluster-galaxy cross-correlation function in the redshift-space on projected scales below 20 Mpc/h, allowing high-fidelity reconstruction of the galaxy infall kinematics~(GIK) as a function of distance to the cluster center. The GIK reconstruction provides a unique probe of the average dynamical mass profile of clusters in the infall region, which will enable stringent tests of the theories of cosmic acceleration when compared with the cluster mass profile measured from weak lensing~\citep{Zu2014}. In addition, the spectr dense spectroscopic sampling of the infall region allows individual measurements of the cluster dynamical mass using the caustics technique.

One of the primary systematics in cluster cosmology is the projection effect due to the 2D aperture adopted by photometric cluster catalogues, leading to the correlation between cluster richness and large-scale overdensity, hence the bias in the large-scale weak lensing signals of clusters~\citep{McClintock2019, Sunayama2023, Salcedo2023}. The JUST cluster infall survey will mitigate this projection effect by adopting a 3D aperture in the velocity phase space for measuring cluster mass observables. Meanwhile, this program will provide a panorama of the star formation, chemical enrichment, and dynamical evolution of galaxies across the cosmic web. Spectral stacking at different cosmic web environments will allow a robust reconstruction of the average histories of star formation and chemical evolution, as galaxies are funneled through the filaments into clusters~\citep{Andrews2013, Lin2022}. By comparing the galaxy
population surrounding clusters with those observed in the non-cluster fields, JUST will provide the key observational evidence on the concept of “nature versus nurture” in galaxy formation. 

\subsubsection{JUST Cluster Galaxy Evolution Survey}

In the nearby universe below $z{<}0.1$, JUST will obtain spectra for galaxies within the virial radius of each SDSS galaxy group~\citep{Yang2007} above $10^{13}\,M_\odot/h$ down to a stellar mass of ${\sim}10^8\,M_\odot$. Focusing on the faint end of the conditional luminosity function of groups~\citep{Lan2016, GM2023}, the JUST cluster galaxy evolution survey will explore the star-forming histories of dwarf galaxies inside the group and cluster-size haloes, and ascertain the existence of a characteristic stellar mass of quenching among the satellites~\citep{Meng2023}. With the accurate measurement of the group/cluster masses, JUST will provide strong constraints on the stellar-to-halo mass relation of the dwarf satellites via abundance matching and satellite weak lensing~\citep{Li2014, Niemiec2017, Sifon2018, Dvornik2020, Danieli2023}. 

The JUST cluster galaxy evolution survey will reveal the co-evolution between cluster galaxies and dark matter haloes, by connecting the spectroscopic observations to the individual halo assembly histories predicted by ELUCID, a state-of-the-art constrained simulation that accurately reconstructed the initial density perturbations within the SDSS volume below $z{=}0.1$~\citep{Wang2014, Wang2016}. Another unique aspect of this program is the exciting synergy with the FAST All Sky HI Survey~(FASHI)~\citep{Zhang2023}, which will provide the largest extragalactic HI catalogue at $z<0.1$ using the Five-hundred-meter Aperture Spherical radio Telescope~\citep[FAST;][]{Nan2011}.

Meanwhile, JUST will reserve a fixed set of fiber assignment for a sample of low-surface brightness targets(e.g., ultra-compact dwarfs) to allow spectral coverage down to ${\simeq}23$ magnitudes per arcsec$^2$ in the r-band~\citep{Liu2020, Wang2023}. For extended sources of interest(e.g., including the outskirts of BCGs and bar galaxies), MOS-mode observations can be supplemented by follow-up observations with the IFU instrument~\citep{Gu2020, Chen2022}. Taking advantage of the synergy with ELUCID and FAST, the versatility of JUST will present an exquisite view of cluster galaxy evolution in the local universe. 

\subsection{Tracing dynamical universe}


The Universe is not static. It is in motion and constantly changing. Time domain astronomy, which focuses on dynamic astronomical events, is a promising method to study this in greater detail. In the 2020 NASA decadal survey for astronomy and astrophysics, it is considered an important research frontier in astronomy \citep{astro2020}. Rapid follow-up observations of unexpected events is crucial in the era of multi-messenger astronomy, allowing astronomers to combine various observation methods such as neutrinos, electromagnetic waves, and gravitational wave signals, which are of great significance for understanding important high-energy astrophysical processes such as black hole and neutron star mergers. The main targets of time-domain astronomy are sporadic events(such as supernova explosions and tidal collapse events.), and the follow-up spectroscopic observation of these events can help to understand the specific physical processes in these transient sources.

There are currently dozens of time-domain astronomical survey projects, such as the Catalina Survey, PanSTARRS, iPTF, ASASSN, ATLAS and ZTF. In the past decade, the number of transient sources discovered has increased tenfold \citep{galyam_etal_13}. The first gravitational wave electromagnetic counterpart was discovered in 2017 \citep{abbott_etal_17} and confirmed to be a millennium nova(kilonova; \citealt{coulte_etal_17}). At present, the number of supernovae discovered is increasing year by year, exceeding one thousand per year. Based on large sample studies, new types of supernovae and explosive physical processes have been discovered. At the same time, new processes of active galactic nucleus explosions and tidal disruption events are also being continuously discovered. Time-domain astronomy has evidently become one of the fastest developing frontier astrophysical research fields. The study of time-domain astronomy can answer the following important questions: What is the explosive process of the evolution of massive stars to their final stages? What are the precursor stars of Type Ia supernovae? How did they erupt? Why does the universe accelerate its expansion? What determines the mass, spin, and radius of a dense star? How do supermassive black holes accrete and grow? Although astronomers have made some progress in addressing these issues, they are still far from fully understanding the physical reasons behind these phenomena.

In the future, surveys like LSST, CSST and WFST will obtain larger transient source samples. It is expected that hundreds or thousands of supernovae and other explosive phenomena will be discovered every night. These future surveys will significantly expand the redshift coverage of transient sources and expand the observation wavelength ranges. Space telescopes such as the Einstein Probe(EP), Swift, and Wide-field Infrared Survey Explorer(WISE) will observe the transient sources in X-ray, ultraviolet, and infrared bands, respectively. It can be expected that these larger samples will bring higher statistical significance, to reveal systematic differences among different types of transient sources and to discover extreme cases in each category. For example, in the past decade, the increasing number of transient source events has spawned research on the relationship between Type la supernovae and the star formation rate in their host galaxies \citep{jones_etal_18}, and the discovery of Type II supernovae that lasted for a year \citep{arcavi_etal_17}, as well as a new type of thermonuclear explosion supernova(SNe Iax; \cite{foley_etal_13, jha_17}).


The photometric detection of supernovae or other transient sources is only the first step. Only by completing the second step of spectral observation can the physical origin of these transient sources be clearly explained. According to Blagorodnova et al.(2018) \cite{blagor_etal_18}, compared to photometric observations, the spectroscopic observation of transient sources is still insufficient. With the development of more photometric surveys, this difference will only become more severe in the future. JUST, with hundreds of optical fibers, can effectively carry out spectroscopic observations of a large number of transient sources by assigning each target with a fiber. It will also provide information on the two-dimensional kinematics and chemical properties of the host galaxy of the transient source with the fibers forming an IFU array, providing first-hand data for studying its triggering environment mechanism.

The transient sources may be induced from many different high-energy phenomena. Among them, events such as gamma-ray bursts, supernovae, and tidal disruption events are generated by cataclysmic processes. AGN flares, X-ray binary bursts, and rapid radio bursts involve periodic and intense physical processes near black holes or compact objects with strong magnetic field. The study of these phenomena not only reveals the specific physical mechanisms, but also helps to test basic theories such as relativity under extreme conditions. JUST will primarily focus on follow-up spectral observations of various transient sources, which is crucial for revealing the driving mechanisms of transient sources.

\subsubsection{Supernova identification and classification}

Important for cosmological research, Type Ia supernovae can serve as standard candles for cosmological distance determination, ultimately leading to the discovery of accelerated expansion of the Universe. High redshift supernovae are mainly discovered through photometric methods, and subsequent spectral analysis helps to distinguish different types of supernovae. On one hand, distinguishing the different types of supernovae can reduce the impact of other types of supernovae on the distance measurement of high redshift galaxies, improving the accuracy of galaxy distance measurement, to better constrain on the accelerated expansion of the universe. On the other hand, analyzing supernova subclasses can help in understanding the basic parameters of precursor stars, the physical processes of explosions, and the interaction between the outflow material and the interstellar medium. 

JUST is capable of rapid response and subsequent spectral observations of supernovae at moderate redshifts($z\sim0.1-0.3$). Within this redshift range, the magnitude of Type Ia supernovae ranges from 18 to 22 magnitudes. The aperture of this telescope is large enough to accomplish this, and the observation conditions at its location(Lenghu) are excellent, which can allow the recording of high signal-to-noise ratio spectra of these sources \citep{2021Natur.596..353D}. This will significantly increase the number of supernova observations in the medium redshift range and may potentially discover new supernova types.

\subsubsection{Gravitational wave electromagnetic counterpart properties}

The discovery of gravitational wave GW150914 was a milestone event in gravitational wave astronomy, which confirmed the existence of a black hole merger for the first time \citep{2016PhRvL.116x1102A}. However, the electromagnetic wave counterpart of the gravitational event was not discovered until 2017, when global synchronous observations of GW170817 confirmed its electromagnetic counterpart for the first time as a binary neutron star merger event \citep{2017ApJ...848L..12A, 2017PhRvL.119p1101A}. Within minutes to hours, Chile’s Swope telescope confirmed an optical flare event in NGC 4993 galaxy. In the following weeks, observatories around the world conducted follow-up observations of the event in different wavelengths, providing a panoramic view of the physical process of the binary neutron star merger event \citep{2017Sci...358.1520C}. The visual magnitude of the optical counterpart of this binary neutron star merger event varies between 17.5 and 23 magnitudes, and JUST can also perform spectral observations of this source and others like it. In the future, more gravitational wave events will be detected. Timely follow-up spectroscopic observation of the source is very important to provide additional information(such as chemical abundance, redshift, and kinematics) to reveal the physical properties of gravitational wave sources.


\subsubsection{The physical process of tidal disruption events}

If a star is too close to a supermassive black hole, it will be disrupted by tidal forces, causing about half of the material to be accreted, resulting in flares at optical, infrared, ultraviolet, X-ray, and other wavelengths. This is known as a tidal disruption event(TDE), which was theoretically proposed in 1970s \citep{1975Natur.255..102H, 1979SvAL....5...16L, 1988Natur.333..523R, 
1989IAUS..136..543P, 1989ApJ...346L..13E, 1999ApJ...514..180U} and observationally confirmed in 1990s \citep{1996A&A...309L..35B, 1999A&A...350L..31G, 1999A&A...349L..45K, 2000A&A...362L..25G}. It has become one of the most important targets in time-domain astronomy. With the advancement of various photometric surveys(such as South Sky LSST and North Sky WFST), a large number of TDEs will be discovered. For example, WFST in China expects to discover tens to hundreds of TDEs annually and to obtain complete light curves, including the early brightening phase. TDEs are one of the main targets of the Einstein Probe X-ray telescope in China. TDE detection is an important method to observe supermassive black holes(including quiescent ones) and provides information on black hole mass and spin, accretion disk physics, strong field gravity, and black hole environment(gas, dust environment, and stellar properties). TDEs are also useful to identify intermediate mass black holes, with the potential to resolve the mass gap between stellar mass black holes and supermassive black holes, completing the evolutionary landscape of black holes.

JUST can efficiently perform follow-up spectroscopic observations of TDEs detected by WFST and EP. Its 4.4-meter aperture, fast pointing adjustment, same observation location, and medium resolution spectrograph make it perfectly compatible with WFST(a 2.5-meter telescope) to carry out joint measurements of TDEs. The typical brightness of TDEs is $\sim20-23$ mag(with a redshift range of $0.3-1$), and the light curve variation period is on the order of months, allowing a considerable success rate in obtaining TDE spectra with redshifts below 1. The acquisition of TDE spectra can provide important information such as accretion disk wind properties, stellar/accretion and disk chemical composition \citep{2018ApJ...869..118D, 2020MNRAS.494.4914P}. Combined with the light profile curves of other photometric surveys, it will significantly improve the understanding of the physical mechanisms of TDEs, as well as the strong gravitational field properties. JUST can also provide key spectroscopic evidence for the tidal disruption of white dwarfs in intermediate mass black holes that have been discovered. 
The IFU observation mode is expected to obtain spectroscopic information on host galaxies, measuring their redshift, dispersion velocity, and chemical composition, to provide further observational constraints on the coevolution of galaxies and supermassive black holes.

\subsubsection{Long term monitoring of active galactic nuclei(reverberation mapping)}

JUST can also monitor the long-term spectral variability of active galactic nuclei(AGN). A considerable number of AGN with redshift less than 1 have $r$-band magnitudes brighter than 22 mag, suitable for future observation with JUST. By analyzing the time delay between the variability of the emission lines and the continuum, the reverberation mapping method can be used to analyze the structural characteristics of the broad line region(BLR) near the black hole, and to estimate the mass of the black hole. In addition, by observing the post spectral variability of some sudden flare phenomena in AGN, we can understand the physical reasons behind the changes in the continuum, broad line structure, and kinematics of AGN with the variation of the accretion rate, to better understand the physical processes of accretion by supermassive black holes.

\subsection{Detection and characterization of exoplanets}


The third category of science motivations for JUST is exoplanet detection and characterization. With its high-resolution spectrometer, JUST will enable the discovery of a substantial number of cold giant planets by employing a combination of radial velocity(RV) and astrometric analyses. In its upgraded phase, JUST will feature an exceptionally high-precision spectrograph designed for detecting Earth-like planets. Leveraging these advanced capabilities, JUST will further enable the characterization of the atmospheres of hot exoplanets, contributing valuable insights into their formation and evolution.

\subsubsection{Detection of cold giants}

The planets in our Solar System and most of the known
exoplanets are thought to form in a bottom-up fashion through
collisions of dust, pebbles, and planetesimals. This
so-called ``core accretion''(CA) mechanism is able to form Jupiter-like 
planets through the processes of core formation, envelope formation
and contraction. However, this formation channel is probably not
efficient to form substellar companions on wider orbits before the dispersion of a
protoplanetary disk in $\sim$10\,Myr \citep{kratter16}. These objects
are more likely to form like stars in a top-down fashion through
the so-called ``gravitational instability''(GI) mechanism
\citep{boss97}. However, due to the flexibility and ambiguity of the
features of substellar companions predicted by CA and GI, it is challenging to determine
which formation channel is responsible for specific giant companions
such as the four directly imaged giant planets around HR 8799
\citep{marois08}. Hence, a statistically significant sample of giant planets on
wide orbits(or cold giants) would be essential to statistically distinguish between GI
and CA and draw a boundary between these two formation channels. 

Thanks to the high precision astrometry catalogs released by Gaia
\citep{gaia16,gaia18,gaia21,gaia23} and the long baseline formed by Gaia and its precursor,
Hipparcos \citep{perryman97,leeuwen07}, many substellar companions
detected by the radial velocity method are confirmed and their absolute masses are determined
by combined analyses of radial velocity(RV) and astrometry data
\citep{snellen18,brandt19,kervella22, feng22}. However, these detection are limited to super-Jupiters or more massive companions due to the limited precision and time span of the current Gaia data and the limited number of stars with high precision RV data. 
While the precision and time span of Gaia data will be significantly improved in Gaia DR4, it is hard to significantly increase the current sample of stars with high precision RVs because of the limited number of high resolution spectrographs and the low efficiency of the current high precision RV survey. 

To facilitate the detection of large number of cold giants with the combined RV and astrometry method, JUST will be equipped with the High Resolution Spectrograph(HRS), which can measure RV with a precision of about 1\,m$s^{-1}$. HRS will be a fiber-fed, white-pupil spectrograph with a design resolution of R=60,000-80,000 and a wavelength of 380-760nm. The instrument design will be based on the successful high resolution spectrograph on the LAMOST \citep{tianyi19} and HARPS-N \citep{cosentino12} on the TNG telescope. In order to obtain a precision radial velocity(PRV), HRS will be enviromentally stabilized in the vacuum enclosure and via two optical fibers will provide simultaneous measurement of the science source and a spectral calibration source. Like other PRV instruments, the HRS will includes three main subsystems, 1) front-end module to correct for atomspheric dispersion, reimage the telescope beam onto the science fiber, stablize the image with fast tip-tilt corrections, 2) calibration unit to enable the injection of different light sources and 3) spectrograph is vibrationally and thermally isolated from the room. To ensure optimal optical performance and superior angular resolution, HRS will be integrated with the first Nasmyth focus of JUST.

\subsubsection{Detection of Earth twins}

One holy grail of exoplanetology is to find the most Earth-like planets. These so-called Earth twins are Earth-sized planets located in the habitable zones of Sun-like stars \citep{kasting93}. These temperate worlds can sustain liquid water on their surface and probably also have other habitable conditions such as plate tectonics, magnetic fields, and stable orbits. The Earth twins are perfect targets for future missions such as LUVOIR, HabEx \citep{luvoir19,gaudi20} and Habitable World Observatory(HWO; \citealt{mamajek23}). 

However, it is challenging to detect Earth twins due to limited instrumental precision and stellar activity. The measurement error of single RVs is typically $>$0.3\,ms$^{-1}$ for second-generation spectrometers such as ESPRESSO on VLT \citep{pepe10}, Maroon-X on Gemini-North \citep{seifahrt22}, NEID on WIYN 3.5m \citep{schwab16}, and the Keck Planet Finder \citep{gibson16}. With advanced data analysis techniques, we are able to detect RV signals as small as 0.3\,ms$^{-1}$ \citep{feng17,faria22}. 

While instruments like ESPRESSO and KPF have achieved an RV precision of sub-m$s^{-1}$ for detecting habitable Earths, stellar activity introduces noise reaching several m$s^{-1}$, surpassing the planetary signal. The challenge in using RV to detect habitable Earths lies in effectively distinguishing this ``red noise'' from the planetary signal, given its time dependence. Advanced noise modeling techniques such as Gaussian processes have been used to mitigate such red noise \citep{haywood14,rajpaul15}. However, these techniques may lead to false negatives due to over-fitting \citep{feng16,ribas18}. 

To mitigate the impact of wavelength-dependent stellar activity noise on radial velocity, traditional methods involve measuring the intensity of spectral lines characterizing stellar magnetic fields to remove the velocity variations linearly correlated with these so-called ``activity indicators'' \citep{dumusque16a,dumusque16b,zechmeister18}. However, different types of stars respond differently to various stellar activity indicators, and the linear removal of velocity correlated with these indicators introduces inherent noise. Therefore, recent research favors directly selecting spectral lines from the spectrum that are less ``contaminated'' by stellar activity \citep{dumusque18, lisogorskyi19}.

In the upgraded phase of JUST instrumentation, an ESPRESSO-like spectrograph, named Extremely high Resolution Spectrograph(ERS), will be built for the detection of Earth twins. ERS will have a resolution of at least 100,000 and can measure RV with a precision of about 0.1\,m$s^{-1}$. It will be built following the design of CHORUS on GTC\footnote{\url{https://www.nao.cas.cn/gtc/hrs/gkoverview/}}.
With this spectrometer, JUST will survey a sample of 20-40 Sun-like nearby stars over 5 years to discover Earth twins. Given the uncertainty of the current occurrence rate of Earth twins \citep{ge22}, we expect to discover at least 1-3 Earth twins as golden samples for future direct imaging missions such as LUVOIR and Habex \citep{luvoir19,gaudi20}.

\subsubsection{Characterization of hot extrasolar giant planets}

One of the primordial goals of exoplanet sciences is to characterize exoplanetary atmospheres and inform the formation and evolution history of the diverse planetary systems \citep{madhusudhan2019}. High-resolution spectroscopy has offered a unique means to measure chemical species in the atmospheres of close-in hot Jupiters because this type of exoplanets so far offers the best signal-to-noise ratio(see a review by \citealp{birkby2018}).  Using the same framework as measuring the extremely precise RV of the planet-hosting stars, this method can be applied to phase-resolved planetary spectral lines which can be identified through the Doppler effects of the orbiting planets. For typical hot Jupiters, the orbital speed is a few orders of magnitude larger than that of the star, and thus the stellar and telluric spectral features are relatively unchanged compared to the planetary spectral lines and can be removed by various detrending methods. The time-varying components of the planetary spectra then reveal compositions in the planetary atmospheres. Typical constituents expected in hot Jupiters' atmospheres include major oxygen- and carbon-bearing species such as ${\rm H_2O}$, ${\rm CH_4}$, ${\rm CO}$, and ${\rm CO_2}$ which are most easily detected in the near-IR and IR wavelengths. In the visible wavelengths, various key heavy elements, including Si, Ti, V, and Fe,  have rich spectral features and have been observed in atmospheres of a dozen hot Jupiters(e.g., \citealp{yan2022}). These refractory elements offer a valuable window to probe the formation and migration history of hot Jupiters \citep{lothringer2021}.

The high-resolution spectroscopy has also been applied to the atmospheric characterization of directly imaged exoplanets, i.e., giant planets that are hot, self-luminous, and with large orbital separations.  Key molecules including CO and ${\rm H_2O}$ have been identified in several directly imaged exoplanets;  isotopes are also within reach for some of the best targets \citep{currie2023}. In addition to composition measurements, the rotationally induced spectral line shapes allow us to determine the rotation periods of directly imaged exoplanets, an important piece of information for tracking how planets accreted their angular momentum when they grew within the disk \citep{snellen2014}. 

The ERS in the upgraded phase of JUST instrumentation should be able to carry out spectroscopic surveys  of dozens of hot Jupiters, yielding statistical trends of metallicity and carbon-to-oxygen ratios for the hot Jupiter population. Equipped with extreme adaptive optics, we expect to characterize several directly imaged exoplanets and measure their atmospheric chemical inventories and spin states.

\section{Summary} \label{sec:summary}

JUST is a 4.4-meter telescope equipped with a segmented primary mirror and a lightweight framework, allowing for reduced construction costs and rapid switching between observation targets. It features two Nasmyth foci, each offering a field of view of 10\,arcmin and 1.2\,degree, with the ability to alternate between them by rotating the tertiary mirror(M3). The telescope also boasts three types of spectrographs: a multiple-fiber medium-resolution spectrometer, an IFU array and/or a long-slit spectrograph, and a multiple-fiber high-resolution spectrometer.

JUST will be installed and operated at a high-quality site with an altitude of 4322 meters on Saishiteng Mountain in Lenghu town, Qinghai province. Expected to achieve first light in 2026, it is poised to become the most powerful telescope for spectroscopic observations  in China for a considerable period. Upon completion, JUST will focus on research in three main directions:(1) Exploring the dark universe through spectroscopic surveys of numerous galaxies in the cosmic web;(2) Tracking the dynamic universe by conducting follow-up spectroscopic observations of various transient sources; (3) Detecting and characterizing exoplanets through the acquisition of high-resolution stellar spectra and the precise measurement of sub-ms$^{-1}$ RV. The JUST project is anticipated to produce impactful research outcomes in the fields of dark matter, dark energy, transient astronomy, and exoplanet searches.

\begin{acknowledgements}

We thank Shanghai Jiao Tong University for supports of building the JUST telescope, Qinghai provincial government and Haixi prefecture for supports on providing the site, dome and infrastructure.
This work is supported by  “the Fundamental Research Funds for the Central Universities”, 111 project No. B20019, and Shanghai Natural Science Foundation, grant No.19ZR1466800. 

\end{acknowledgements}



\bibliographystyle{ati} 
\bibliography{References}      

\label{lastpage}

\end{document}